\documentclass[%
 aip, apl,
 amsmath, amssymb,
 reprint,%
floatfix
]{revtex4-1}

\usepackage{graphicx}
\usepackage{dcolumn}
\usepackage{bm}

\usepackage[utf8]{inputenc}
\usepackage[T1]{fontenc}
\usepackage{mathptmx}
\usepackage{etoolbox}

\makeatletter
\def\@email#1#2{%
 \endgroup
 \patchcmd{\titleblock@produce}
  {\frontmatter@RRAPformat}
  {\frontmatter@RRAPformat{\produce@RRAP{*#1\href{mailto:#2}{#2}}}\frontmatter@RRAPformat}
  {}{}
}%
\makeatother
\begin{document}

\title[draft]{Scaling NbTiN-based ac-powered Josephson digital to 400M\,devices/cm$^2$}
\author{Anna Herr}
 \email{anna.herr@imec-int.com}
 \affiliation{imec USA, Kissimmee, FL 34744, USA}
\author{Quentin Herr}
 \affiliation{imec USA, Kissimmee, FL 34744, USA}
\author{Steve Brebels}
 \affiliation{imec, 3001 Leuven, Belgium}
\author{Min-Soo Kim}
 \affiliation{imec, 3001 Leuven, Belgium}
\author{Ankit Pokhrel}
 \affiliation{imec, 3001 Leuven, Belgium}
\author{Blake Hodges}
 \affiliation{imec USA, Kissimmee, FL 34744, USA}
\author{Trent Josephsen}
 \affiliation{imec USA, Kissimmee, FL 34744, USA}
\author{Sabine ONeal}
 \affiliation{imec USA, Kissimmee, FL 34744, USA}
\author{Ruiheng Bai}
 \affiliation{Laboratory of Atomic and Solid State Physics, Cornell University, Ithaca, NY 14853, USA}
\author{Alexander Jarjour}
 \affiliation{Laboratory of Atomic and Solid State Physics, Cornell University, Ithaca, NY 14853, USA}
\author{Katja Nowack}
 \affiliation{Laboratory of Atomic and Solid State Physics, Cornell University, Ithaca, NY 14853, USA}
\author{Anne-Marie Valente-Feliciano}
 \affiliation{Thomas Jefferson National Accelerator Facility, Newport News, VA 23606, USA}
\author{Zsolt T\"{o}kei}
 \affiliation{imec, 3001 Leuven, Belgium}

\date{\today}

\begin{abstract}
We describe a fabrication stackup for digital logic with 16
superconducting NbTiN layers, self-shunted $\alpha$-silicon barrier
Josephson Junctions (JJs), and low loss, high-$\kappa$ tunable
hafnium–zirconium oxide (HZO) capacitors. The stack enables
400\,MJJ/cm$^2$ device density, efficient routing, and AC power
distribution on a resonant network. The materials scale beyond 28\,nm
lithography and are compatible with standard high-temperature CMOS
processes. We report initial results for two-metal layer NbTiN wires
with 50\,nm critical dimension. A semi-damascene wire-and-via process
module using 193i lithography and 50\,nm critical dimension has shown
cross-section uniformity of $1\%/1\sigma$ across the 300\,mm wafer,
critical temperature of 12.5\,K, and critical current of 0.1\,mA at
4.2\,K. We also present a design of the resonant AC power network
enabled by NbTiN wires and HZO MIM capacitors. The design matches the
device density and provides a 30\,GHz clock with estimated efficiency
of up to 90\%. Finally, magnetic imaging of patterned NbTiN ground
planes shows low intrinsic defectivity and consistent trapping of
vortices in 0.5\,$\mu$m holes spaced on a 20$\times$20\,$\mu$m$^2$
grid.
\end{abstract}

\maketitle

\section{Introduction}
Unsustainable demand for compute power and unsustainable production
hardware cost \cite{strubell2020energy, li2020chiplet} open the door
to new technology in the post-Moore era. Superconducting digital logic
has the potential to provide a sustainable solution for
large-scale compute applications positioned between mature CMOS and
long-horizon quantum computing. The differentiators for
superconducting digital are energy efficiency, high computational
density, and high interconnect bandwidth. These features uniquely
enable real-time AI training models and greater inclusiveness by
distributing to edge systems compute power that currently must be
deployed in a centralized data center. Development of a
superconducting ecosystem would enable this market and would also
enable tactical applications in communications
\cite{ilderem2020technology} and signal processing
\cite{ayala2021scalability}, and would lead to further innovations in
areas ranging from quantum computing \cite{mcdermott2018quantum} to
reversible logic \cite{wustmann2020reversible} and neuromorphic
computing \cite{kuncic2021neuromorphic}.

While superconducting integrated circuits have existed for decades at
a certain modest device count, success in scaling further has met with
challenges. Three major limitations include feature size and layer
count in the fabrication process, power distribution to current-bias
the Josephson junctions, and superconducting flux trapping
failures. In this paper we report progress in all of these areas,
centered around a fabrication stack with the required features enabled
by an updated materials set.

All state-of-art superconducting fabrication processes for digital
circuits are Nb-based \cite{yohannes2005characterization,
  berkley2010scalable, tolpygo2014fabrication, egan2022true}. As an
elemental superconductor, Nb has a relatively high critical
temperature of 9.2\,K and can be fabricated using conventional DC
sputtering and etch.  However, processes scaled to an increased number
of metal layers and feature sizes down to 0.25\,$\mu$m have run into
fundamental limitations. Problems include 1) unstable material
properties caused by a low diffusion constant and complex oxidation,
affecting tolerances at sub-micron dimensions \cite{herr2018yield,
  pinto2018dimensional, gupta2016preserving, gubin2005dependence,
  tolpygo2021inductance} 2) low inductance per unit length
\cite{tolpygo2014inductance, tolpygo2020increasing} limiting routing
density and signal integrity, 3) microwave losses due to inherent
materials properties and an easily contaminated surface
\cite{yogi1981microwave}, and 4) a low processing temperature
requirement \cite{crauste2013effect}. Nb exhibits degradation of the
superconducting properties top-to-bottom through the stack due to the
cumulative effects of fabrication processing
\cite{verjauw2021investigation}. Wire critical temperature
distributions vary widely from process-to-process and from
wafer-to-wafer. Further degradation due to hydrogen poisoning is
reported \cite{amparo2010investigation}. The low processing
temperature is incompatible with standard industry dielectric process
\cite{lindemann2007selective}, with the overall effect limiting
feature size, layer count, and process control.

Fabrication of Nb/AlOx/Nb Josephson Junctions (JJs) is accessible to both
academia and industry labs, but these junctions are at the limit of
scalability as 1) the barrier is too thin to scale to higher critical
current densities without degradation of junction quality resulting in
increased losses and spreads \cite{tolpygo2017properties}, 2) the
capacitance of the barrier limits device speed \cite{
likharev1985introduction}, and 3) the low thermal budget of
150-200$^{\circ}$\,C \cite{migacz2003thermal} compromises subsequent fabrication
processing of the backend. Integration density has not exceeded
1\,MJJ/cm$^2$ with only four layers of metal \cite{semenov2017ac}, and
has not exceeded 100\,kJJ/cm$^2$ with eight
layers \cite{herr2015reproducible}. Published approaches for further
increase in integration density are based on “self-shunted” junctions
with lossy barriers. Lossy barriers compromise circuit performance by
increasing fabrication spread in the case of high-Jc ultra-thin
barrier AlOx junctions \cite{tolpygo2016superconductor}, or by
reducing speed in the case of Nb-doped Si
junctions \cite{olaya2010digital, gudkov2012properties}.

Current fabrication technology does not support efficient
implementation of AC power distribution. An industry shift from DC
power to resonant AC was a major advance for energy efficiency,
scalability, timing, and gate-delay of superconducting digital
circuits \cite{farrell2018superconducting, vesely2018pipelined,
  egan2022synchronous, semenov2017ac, ayala2020mana}. However, the
most advanced AC resonator networks are limited in density and
power-efficiency due to the lack of critical fabrication features. The
geometric transformer used at every bias tap limits integration scale
to about 4\,MJJs/cm$^2$. Further miniaturization of the transformer
increases the input power and on-chip dissipation. Proposals for
capacitive coupling
\cite{strong2020capacitive, herr2020capacitively} would enable
higher integration scale but requires small, high-$\kappa$ MIM
capacitors. Static power dissipation in the resonator may dominate the
total power budget, arising from losses in the Nb wires and
dielectrics. AC power distribution across multiple chips is
constrained by a lack of tunability of the high-Q resonant power
network, needed to compensate for fabrication-induced resonance
frequency spreads.

The largest SFQ circuits have been powered by low-speed meander lines
\cite{herr2015reproducible, semenov2017ac}, but meander lines at GHz
frequencies have shown limitations due to reflections of the traveling
waves in the package, speed-of-light uncertainty on-chip, and power
dissipation \cite{herr20138}. A meander-line 4-bit AQFP CPU was
demonstrated at 100\,kHz, and subcomponents at 2.5\,GHz. A resonant
clock network based on standing waves was first put forward in the
context of QFPs \cite{hosoya1991quantum} and was finally demonstrated
in RQL \cite{strong2022resonant}, with small circuits operating at
10\,GHz and circuits of 40,000\,JJs at 3\,GHz. Further scaling was
limited by the large physical size of the transformer in each bias
tap. Scaling of superconductor ICs requires fundamental changes to
move beyond all of these limitations.

\section{Fabrication Stack}
\begin{figure}
\includegraphics[width=3.3in]{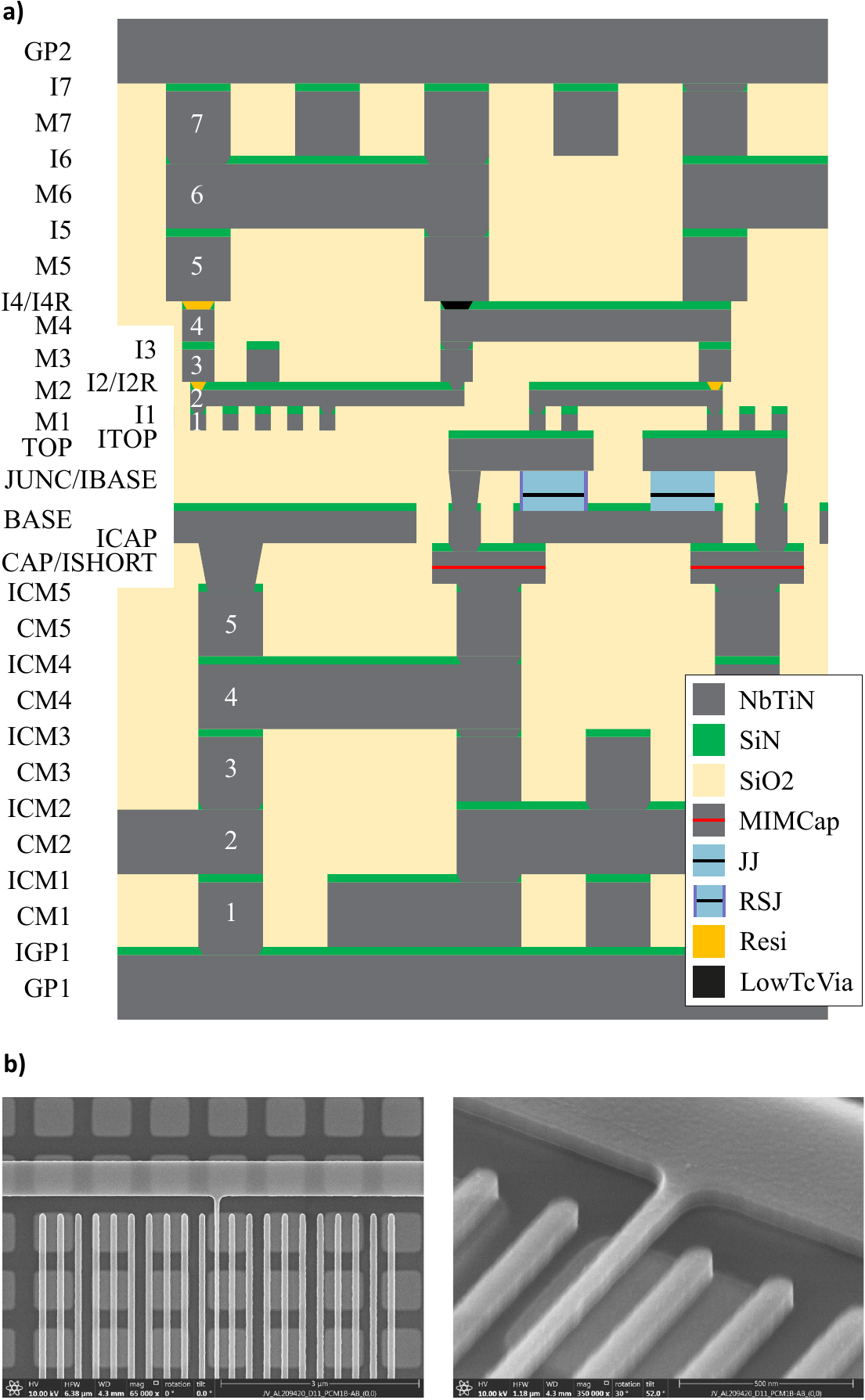}
\caption{\label{fig1} Fabrication stack. a) The proposed fabrication
  stack has 16 metal with power distribution layers on the bottom,
  tunable capacitor and JJ at the center, and wireup layers on top. b)
  SEM images of two metal layers with 50\,nm critical dimension.}
\end{figure}

We propose three process modules: NbTiN
high-kinetic-inductance wires, Nb/Si/Nb Josephson junctions with
self-aligned shunts, and tunable HZO MIM capacitors. All of these
materials have been explored and developed in other contexts, but have
not been applied to digital integrated circuits. Fig. ~\ref{fig1}
shows our proposqed fabrication stack with 16 metal layers and two
ground planes dedicated to flux-trapping mitigation. Current
development is at the level of two metal layers. The stack has
balanced resources supporting an integration of 400\,MJJs/cm$^2$
matched to HZO-capacitor bias taps, 6 NbTiN backend routing layers
with resistors implemented as vias, and 4 NbTiN layers for resonant
power distribution. The high kinetic inductance of the wiring supports
the high density of junctions and bias taps. In a companion paper
\cite{herr2023superconducting}, we present circuit designs and
projected performance of logic and memory mapped onto the fabrication
stack.

The frontend of the stack is based on $\alpha$-Si barrier
junctions with high quality factor, small capacitance, and integrated
shunts. Such junctions meet circuit performance and density
requirements: high critical current, high quality, low capacitance,
high reproducibility, and tolerance of high temperature
processing. These junctions were introduced decades ago
\cite{kroger1979niobium},\cite{kroger1985hydrogenated} with impressive
characteristics and have recently come back into use
\cite{baek2006co}. Relative to Nb/AlOx/Nb, Si-based junctions have
10$\times$ the critical current density at 4$\times$ thicker barrier.
Critical current density of 100\,$\mu$A/$\mu$m$^2$ produces a
35\,$\mu$A minimum junctions with a diameter of 0.21\,$\mu$m. The
small capacitance of these junctions enables them to reach the
fundamental switching speed of 1.1\,ps based on a gap of 1.8\,mV,
resulting in 3$\times$ reduced gate latency. At the 193i lithography
node the total area of each junction instance is less than
0.25\,$\mu$m$^2$ including the via to the base electrode and
integrated shunts. A resistive passivation with $\rho\approx 1
\text{m}\Omega\text{cm}$ can act as a self-aligned donut shunt
resistor.

The backend process is based on NbTiN wires recently introduced by the
quantum and RF superconducting communities driven by the need for
small dimensions and extremely low losses \cite{shan2016parametric,
  valente2015growth, mazin2020superconducting,
  hahnle2021superconducting}.  Stable and controllable material
parameters are a fundamental advantage of this choice. Relative to Nb,
NbTiN has up to a 1.8$\times$ higher critical temperature at 17.3\,K,
and a 1.8$\times$ higher gap voltage at 5.2\,mV. Ti is a nitrogen
getter, so higher Ti composition produces a lower number of vacancies
and high stability, with processing temperatures up to 1,000\,C
\cite{valente2015growth, hahnle2021superconducting,
  valente2020material, shu2021nonlinearity}.  NbTiN interconnects have
extremely low microwave surface resistance \cite{swails2018depairing,
  cyberey2019growth} that translates into a 10$\times$ higher
throughput and 10$\times$ less energy per bit compared to Nb. Reported
NbTiN wires exhibit stable superconducting properties down to a
10$\times$10\,nm$^2$ cross section with a critical temperature up to
15\,K and a critical current density of 140-200\,mA/$\mu$m$^2$ (see
\cite{swails2018depairing} and references thereof). NbTiN has a
4$\times$ lower resistivity translating into a 2$\times$ higher
critical current than NbN \cite{clem2012kinetic}.

The high kinetic inductance of NbTiN enables physically small,
fixed-inductance-target interconnect without meanders, and upholds
signal integrity at deep-submicron by minimizing parasitic mutual
inductance arising from the geometric inductance. Target inductance
values have low sensitivity to the placement of ground planes and
interlayer dielectric (ILD) thickness. A central feature is the
ability to design routes to fixed-inductance targets through
utilization of layers with different cross-sections that produce a
10$\times$ spread in inductance per unit length. Three
pairs of routing layers, labeled M1-M6 in Fig.~\ref{fig1}, with fine
features (50\,nm) on the bottom of the stack and coarse features
(200\,nm) on top, are modeled after CMOS. This methodology would be
suitable for inductive and PTL routes using commercial place and route
algorithms.

Fabrication of high-kinetic-inductance NbTiN wires requires tight
control on wire cross-section that is consistent with 193i lithography
and 300\,mm wafers.  Recent progress on a NbTiN backend process
reported by imec \cite{pokhrel2023towards} shows 50\,nm
critical-dimension (CD) wires with cross-wafer CD one-sigma of 1\%. The
process follows imec's advances in semi-damascene Ru with 12\,nm CD
for advanced-node CMOS, with cross-section one-sigma of 1\% at 12\,nm
feature size \cite{murdoch2020semidamascene}. Semi-damascene achieves
high yield and tight cross-section control by wire etch and
low-aspect-ratio via fill. Two layers of patterned NbTiN, shown in
Fig~\ref{fig1}b, have a measured critical temperature of 12.5\,K and
critical current density of 8\,mA/$\mu\text{m}^2$, consistent with a
film resistivity of 160\,$\mu\Omega\text{cm}$ and penetration depth of
380\,nm \cite{bartolf2015fluctuation}. Tested samples had film
thicknesses of 50 \& 100\,nm, and wire widths of 50-250\,nm. Structures
up to 100\,$\mu$m show no degradation of superconducting properties as a
function of length.

NbTiN wires and high-$\kappa$ HZO MIM capacitors form the basis for
low-loss, tunable resonant AC power distribution. HZO has been
extensively studied in the context of FeFETs and MMICs
\cite{schenk2013strontium, aldrigo2020tunable}. The capacitors can be
fabricated using established ALD processes \cite{choi2011development}
to support high dielectric constants up to 38, high reproducibility,
and tolerance of high temperature processing. Tunable HZO capacitors
with 10\% tunability with applied voltage have recently been reported
for room temperature applications in logic, memory, and MMICs
\cite{lin2020ferroelectric, woo2021device, yurchuk2014impact}.
Tunability with DC voltage is the key to practical high-Q designs, and
does not create additional energy dissipation. Losses in the resonator
will be dominated by the interlayer dielectric and capacitors, with a
negligible contribution from the NbTiN wires. Low-loss $\alpha$-Si
dielectric has been reported \cite{shu2021nonlinearity} in resonators
with $\tan(\delta) < 10^{-5}$. Experiments are required to determine
and optimize capacitor losses at cryo temperatures. Based on results
for metallic oxides in quantum computing applications
\cite{mcrae2020materials}, we project a resonator quality factor of
$Q=10^5$, which represents a 100$\times$ improvement over SiO2
capacitors for Nb-based superconductor processes.

\section{AC Power Distribution}
\begin{figure*}
\includegraphics[width=7.0in]{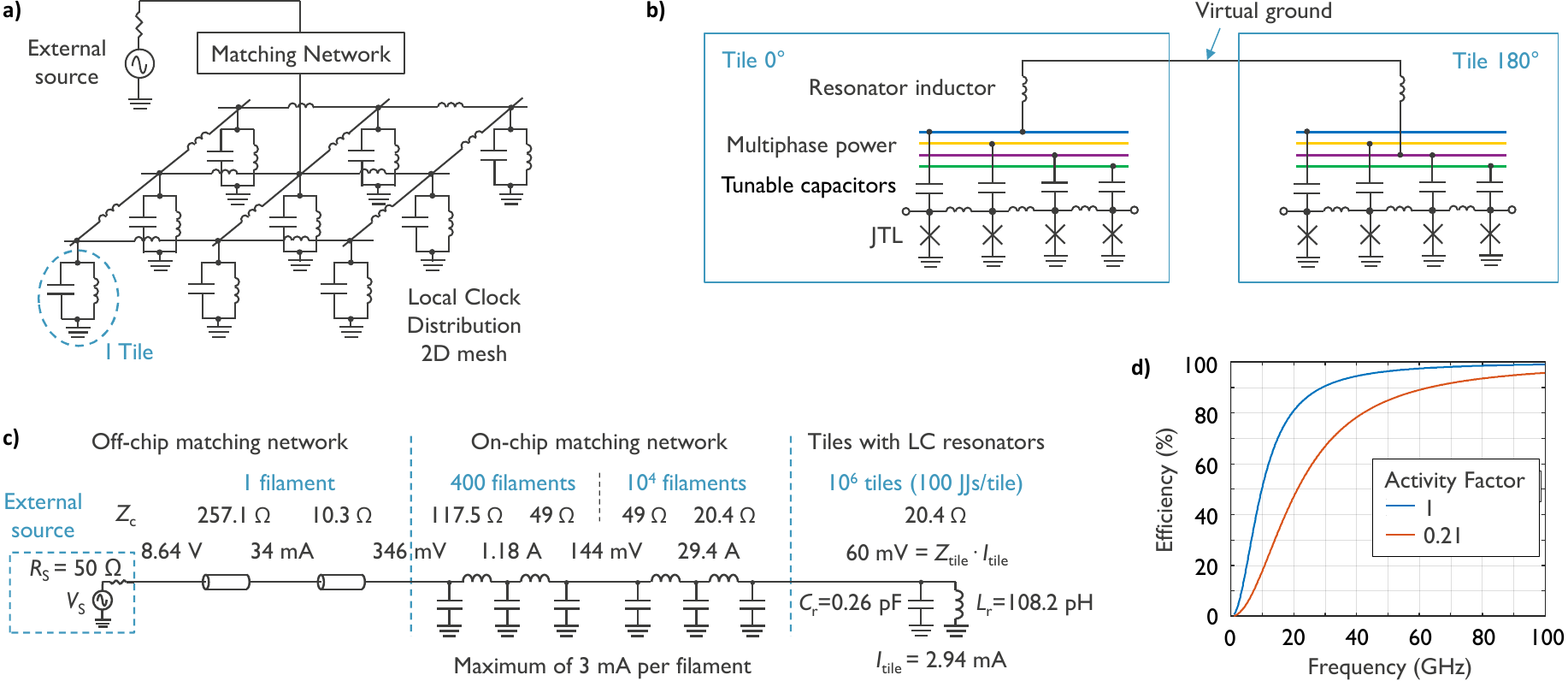}
\caption{\label{fig3} Resonant clock distribution network. a) A
  two-dimensional array of LC resonators coupled through an inductive
  mesh supports a zeroth-order resonance of uniform amplitude and
  phase. Each LC represents a single inductor and multiple capacitors
  to power 100 JJs in one 5$\times$5\,$\mu$m$^2$ tile at the 30\,GHz
  resonance. b) Each JJ in the circuit is connected to one phase from
  the multiphase power lines. Resonator inductors of opposite phase
  are connected together to create a virtual ground. This
  configuration allows connection of a DC bias over the tunable
  capacitors to adjust the resonance frequency. c) A matching network
  provides impedance transformation from 20\,$\mu\Omega$ impedance of
  one resonator phase to 50\,$\Omega$. The impedance of each stage of
  the matching network is governed by the overall design requirement
  that the loaded quality factor be about 100. d) The efficiency of
  the power distribution network as function of resonant frequency
  plotted for different JJ activity factors.
}
\end{figure*}

The power layers of the fabrication stack are designed for a high
integration density implementation of the AC power network with a 2D
array of tightly-coupled local LC resonators as shown in
Fig.~\ref{fig3}a. The design balances multiple features including
density of bias taps, efficient use of metal layers, energy
efficiency, clean resonance mode structure, and tunable frequency. The
design has a 30\,GHz clock frequency and 400\,M bias taps per
1$\times$1\,cm$^2$.  Density is achieved by shunting the JJs with
lumped LC resonators as shown in Fig.~\ref{fig3}b. A critical design
feature is that inductance scales inversely with number of Josephson
junctions. The inductance of a combined 100 taps is 10,000 times
smaller than that of 100 individual taps. In the proposed stack, a
5$\times$5\,$\mu\text{m}^2$ area accommodates 100 JJs with average
critical current of 42\,$\mu$A and a corresponding 100 HZO MIM
capacitors with an average 2.6\,fF.  The maximum number of the JJs per
tile is set by the bias current through the inductor, which must be
within the critical current of the high kinetic inductance wire that
scales with wire cross-section. The NbTiN wire with cross-section
200\,nm$\times$200\,nm has a critical current up to 5\,mA. This
constraint sets the number of junctions per tile.

Scaling the design to 30\,GHz and large chip area of
1$\times$1\,cm$^2$ is achieved by coupling LC tiles with a 2D
inductive mesh. This ideal network presents a zero-order mode
resonance with uniform amplitude and phase. The LC resonance
frequencies however changes over larger chip areas due to variation in
materials and processes. Tuneable HZO capacitors are therefore
introduced to improve homogeneity of the resonance frequency
across-chip. Capacitance is tuned over 10\% by adjusting the DC bias
across the capacitor.

Practical aspects of the design include the feed network, illustrated
in Fig.~\ref{fig3}c, that transforms impedance and stabilizes the
desired zero-order mode. The impedance of an individual LC resonator
tile is two orders of magnitude higher than the characteristic
impedance of the biased JJs, but the impedance of 4\,M tiles in
parallel is still much lower than the 50\,$\Omega$ source.
Implementation of the feed network is based on the same lumped-LC
layout style as the primary resonators, with a progressive decrease in
the number of tiles per stage in the feed network, producing the
corresponding increase in impedance. In this way a hierarchical
structure of LC resonators is formed.  The number of connections
between layers is optimized to cancel parasitic resonances by
providing a regular grid of feed points, and to keep current at any
point of the network below 3\,mA. Inductance of the 2D mesh is
invariant between all layers of the feed network. Capacitance at the
highest level is dominated by mesh-based parasitics to the ground,
which the design must take into account.

Another constraint on the power distribution is amplitude stability
across variable JJ switching activity. High switching activity loads
the network and reduces amplitude. The variation can be expressed as
\begin{equation}
  I_\text{JJ}^\text{max}/I_\text{JJ}^\text{min} \approx 1+Q_\text{ext}/Q_\text{JJ},
\end{equation}
with $Q_\text{ext}$ the quality factor of the external 50\,$\Omega$
source at room temperature and $Q_\text{JJ}$ the quality factor of the
JJ dynamic switching. Stability is guaranteed when the external
quality factor of the resonator is sufficiently low compared to the
quality factor of dynamic switching. The desired external quality
factor is achieved in the design of the matching network.

Not all AC power from the LC resonators is consumed by switching
JJs. Part of it is lost inside the capacitor and inductor of the
resonator. Well below the critical temperature, the loss in the NbTiN
inductor is so low \cite{mazin2020superconducting} that only the loss
in the capacitpr dielectric is significant. The quality factor of the
material loss is therefore $Q_\text{mat} \approx
Q_\text{diel}=1/\tan{\delta}$, the inverse of the loss tangent of the
capacitor dielectric. The useful power consumed by the switching JJs
has an associated quality factor $Q_\text{JJ}=I_\text{JJ}^2/(\alpha
4\pi f_r^2 C_r E_\text{JJ})$ with $\alpha$ the activity factor and
$f_r$ the clock frequency. $C_r$, $I_\text{JJ}$ and $E_\text{JJ}$ are
parameters for a 5$\times$5\,$\mu$m$^2$ tile of 100 switching JJs:
$C_r$ is the resonator capacitance for one tile, $I_\text{JJ}$ and
$E_\text{JJ}$ are respectively the total bias current and total
switching energy for 100 JJs in one tile. The efficiency of the power
distribution is
\begin{equation}
  \eta=\left(1+\frac{Q_\text{JJ}}{Q_\text{mat}}\right)^{-1} \approx
  \left(1+\frac{1}{\alpha}\frac{I_\text{JJ}^2
  \tan{\delta}}{4\pi f_r^2 C_r E_\text{JJ}}\right)^{-1}.
\end{equation}
The efficiency is dependent on the activity factor and improves with
clock frequency as shown in Fig.~\ref{fig3}d. The dynamic power
consumption of the JJs is determined with a bias current $I_\text{JJ}$
of 2.94\,mA and energy consumption $E_\text{JJ}$ of 2.87\,aJ for 100
JJs in one tile. A tile capacitor of 0.26\,pF with a loss tangent of
10$^{-4}$ is used in the efficiency calculation. At 30\,GHz,
efficiency of 90\% is achieved at a high activity factor of 1, and
65\% is achieved at an activity factor of 0.21 that corresponds to the
practical case in efficient logic designs.

\begin{figure}[!ht]
\includegraphics[width=3.4in]{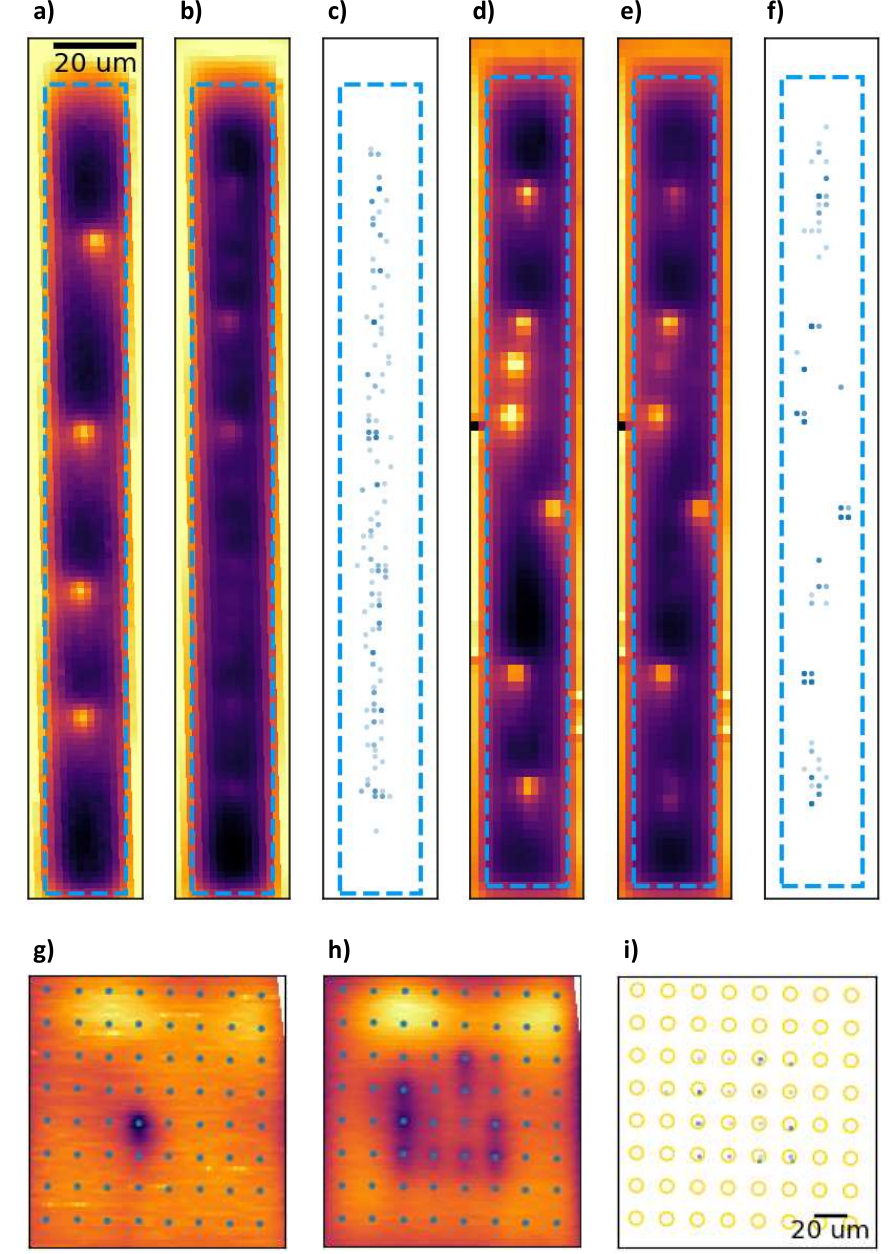}
\caption{\label{fig2} Vortex location in field-cooled NbTiN patterned
  films. a) An image of a 20\,$\mu$m strip fabricated from a 100\,nm
  thick film cooled in 12\,$\mu$T field. The blue box indicates the
  perimeter of the strip. b) Average of images taken over 50 thermal
  cycles. Preferred trapping sites are situated near the center line
  of the strip with a trapping probability of up to 42\% per site. c)
  Extracted locations of vortices observed in the 50 images. Every
  location observed in any of the 50 images is shown as a
  semi-transparent dot. Darker dots indicate that multiple images show
  vortices at these locations. d)-f) A similar progression for a
  50\,nm thick strip field cooled in 14\,$\mu$T field. Preferred
  trapping sites here include locations near the edge of the
  strip. Trapping probability is up to 100\% per site. g) An image of
  a 180$\times$180\,$\mu$m$^2$ ground plane from the same 50\,nm thick
  wafer, patterned with 0.5\,$\mu$m holes spaced 20\,$\mu$m apart. A
  low ambient field resulted in one trapped flux. The edges of the
  ground plane are just outside the imaging area. Vortex locations are
  indicated by the blue dots, extracted from a separate image taken at
  the matching field of 1\,$\Phi_0$ per hole. h) Average of images
  taken across 15 thermal cycles with low fields resulting in 1-4
  trapped flux. i) Extracted locations of vortices observed in the 15
  images. Yellow circles indicate hole locations. Each vortex in the
  15 images is indicated by a semi-transparent blue dot. All vortices
  not expelled from the perimeter of the ground plane reside in the
  holes.}
\end{figure}

\section{Flux Trapping}
Advanced materials and fabrication development would be incomplete
without characterization of flux trapping, which causes variability in
the circuit cooldown to cooldown. This has historically been a
dominant failure mechanism. Pearl vortices form in the ground plane
during cooldown through the superconducting transition temperature
\cite{pearl1964current}, producing parasitic-flux coupling to the
active circuit. The parasitic coupling is up to one-half of an SFQ,
$\Phi_0/2$, if the vortex is located directly under the wire.

Ginzburg-Landau theory \cite{tinkham2004introduction} describes the
physics of vortex motion. Two characteristic energies and two material
constants determine vortex dynamics: magnetic field energy that
depends on London penetration depth, and pinning energy that depends
on coherence length. In practice, defectivity of the material affects
pinning energy and viscosity of the vortex \cite{dhakal2020flux}. The
potential of the vortex can be calculated analytically in simple cases
\cite{brandt2006vortices, shmidt1974vortices} and can be simulated in
complex geometries \cite{kartsev2018opencl, blair2021simulations,
  sadovskyy2015stable}.  The materials aspect of flux trapping has
been studied in the cavity and high current communities, as unpinned
vortices are a major source of loss and reduced critical current
\cite{liarte2018vortex, giroud1992resistive}. These communities have
put extensive effort into material optimization and characterization
of pinning parameters.

The effort in the digital community has been on design of ground plane
voids, or moats, to attract and sequester the flux. Moat design has
been treated empirically through the scanning squid microscopy
measurements \cite{kirtley2010fundamental} of different geometries
\cite{bermon1983moat, jeffery1995magnetic}. Historically, moat design
was not fully effective simply because the moats were spaced too far
apart. A conceptual breakthrough occurred when patterned strips of Nb
less than 20\,$\mu$m wide were shown to completely expel flux
\cite{stan2004critical}. In the context of ICs, this translates to
long moats spaced less than 20\,$\mu$m apart, which have been
effective for large, linear circuits \cite{herr2015reproducible,
  semenov2017ac}. Long moats are incompatible with X-Y routing in
large circuits. Instead, a regular grid of smaller moats has been
adopted with established design rules for moat coupling
\cite{fujiwara2009research, herr2020superconductor}.  Alternately,
design tools \cite{fourie2019coldflux} and modeling
\cite{semenov2016moats} have been proposed to evaluate ad-hoc moat
placement, but these tools consider only the magnetic energy of the
vortex, neglecting material defects.

While mitigations have improved, allowing circuit scale and complexity
to increase, the best understandings of parasitic flux mitigation have
not been fully implemented due to limitations imposed by materials and
fabrication.  Flux mitigation must start with the fabrication stack
including material stability and control, enabling dedicated ground
planes with a relatively high critical temperature
\cite{semenov2016moats}, electrically isolated from the rest of the
circuit \cite{herr2020superconductor}. The miniaturization that comes
with advanced fabrication nodes would only improve flux performance by
placing the moats closer together. The fabrication stack described
here answers to all the above. NbTiN ground planes top and bottom have
critical temperature that can be engineered with material composition.

Fig.~\ref{fig2} shows flux images of the NbTiN ground plane fabricated
in the same process as the wires and patterned into either strips or a
grid of holes. Images can be directly compared to similar experiments
done with Nb \cite{stan2004critical, grigorenko2001direct,
  chiaro2016dielectric}. The primary figure of merit in these
experiments is the critical field, $B_m$ corresponding to onset of the
first vortex trapped in the ground plane. The critical field as
function of strip width is a direct measure of the material parameters
of the film \cite{stan2004critical, clem2012kinetic,
  likharev1971formation}. The critical field of the 10\,$\mu$m NbTiN
strip, $B_m^\text{NbTiN} \approx 40\,\mu$T, is less than the published
value for Nb, $B_m^\text{Nb} \approx 60\mu$T for a film thickness of
210\,nm. The measured critical field for NbTiN depends on the film
thickness, 45\,$\mu$T for 50\,nm and 40\,$\mu$T for 100\,nm. The
decrease in critical field relative to Nb follows an increase of the
Ginsburg-Landau constant $\kappa \approx 80$ for a coherence length of
5\,nm \cite{valente2015growth} and penetration depth of 380\,nm
estimated from the normal-state resistivity and superconducting
critical temperature of the films \cite{bartolf2015fluctuation}.

The NbTiN images in Fig.~\ref{fig2} show that internal defectivity of
the material is quite low. Internal pinning of the material is
characterized by distances from a trapped vortex to the centerline and
to the edge of the strip. The position of the vortices in the 100\,nm
film are near the centerline and there are no consistent pinning
centers. The 50\,nm film images do show consistent defects, including
one that is close to the edge of the film. The difference between the
two films might be explained by columnar growth of NbTiN film with
grain boundaries averaging out with an increase of the
thickness. Finally, the patterned ground plane with 0.5\,$\mu$m holes
spaced 20\,$\mu$m apart shows consistent pinning in the holes at an
applied field up to 240\,nT, typical of digital circuits in a
passively shielded environment. The hole size and spacing is
consistent with high integration density of the stack consuming less
than 1\% of the total area on the chip and allowing for dense routing
with minimal blockages.

\section{Conclusion}
We have proposed a superconductor integrated circuit fabrication stack
with a materials set that provides a foundation for 400\,MJJ/cm$^2$
integration scale in digital superconducting circuits. Fabrication
processes for NbTiN wireup, low-capacitance $\alpha$-silicon barrier
JJs, and low loss, high-$\kappa$ tunable HZO capacitors are compatible
with 300\,mm CMOS processes and thermal budgets. The stack uses
standard inter-layer dielectric processes and 28\,nm
lithography. Initial fabrication and test cycles have shown wires with
properties meeting the design requirements and good flux trapping
properties sufficient for large scale integrated circuits. Further
development is progressing with Josephson junctions and MIM
capacitors. The fabrication stack has been codeveloped with a resonant
clock network scalable for 400M taps in a chip area of
1$\times$1\,cm$^2$ and clock frequency of 30\,GHz. In a companion
paper \cite{herr2023superconducting} we present circuit designs for
logic and memory built on this foundation.

\begin{acknowledgments}
Work at imec and imec USA is supported by imec INVEST+ and by Osceola
County. Work at Cornell is supported by the Cornell Center for
Materials Research with funding from the NSF MRSEC program
(DMR-1719875). Work at Jefferson Laboratory is supported by the
U.S. Department of Energy, Office of Science, and Office of Nuclear
Physics under Contract No. DE-AC05-06OR23177.
\end{acknowledgments}


\section*{References}
\nocite{}
\bibliography{compendium}

\end{document}